\documentstyle[prl,aps,twocolumn]{revtex}
\begin{document}
\draft

\twocolumn[\hsize\textwidth\columnwidth\hsize\csname @twocolumnfalse\endcsname
\title{Quantum Nucleardynamics as an $SU(2)_N \times U(1)_Z$ Gauge Theory}
\author{Heui-Seol Roh\thanks{e-mail: hroh@nature.skku.ac.kr}}
\address{BK21 Physics Research Division, Department of Physics, Sung Kyun Kwan University, Suwon 440-746, Republic of Korea}
\date{\today}
\maketitle

\begin{abstract}
It is shown that quantum nucleardynamics (QND) as an $SU(2)_N \times U(1)_Z$ gauge
theory, which is generated from quantum chromodynamics (QCD) as an $SU(3)_C$ gauge
theory through dynamical spontaneous symmetry breaking, successfully describes nuclear
phenomena at low energies. The proton and neutron assigned as a strong isospin doublet
are identified as a colorspin plus weak isospin doublet. Massive gluon mediates strong
interactions with the effective coupling constant $G_R/\sqrt{2}= g_n^2/8 M_G^2 \approx
10 \ \textup{GeV}^{-2}$ just like Fermi weak constant $G_F/\sqrt{2} = g_w^2/8 M_W^2
\approx10^{-5} \ \textup{GeV}^{-2}$ in the Glashow-Weinberg-Salam model where $g_n$
and $g_w$ are the coupling constants and $M_G$ and $M_W$ are the gauge boson masses.
Explicit evidences such as lifetimes and cross sections of nuclear scattering and
reaction, nuclear matter and charge densities, nucleon-nucleon scattering, magnetic
dipole moment, gamma decay, etc. are shown in support of QND. The baryon number
conservation is the consequence of the $U(1)_Z$ gauge theory and the proton number
conservation is the consequence of the $U(1)_f$ gauge theory.
\end{abstract}

 ] \narrowtext

There are, in the one hand, two distinct problems in quantum chromodynamics (QCD)
\cite{Frit} with quarks and gluons as fundamental constituents. One is the
confinement, which is not rigorously explained in the low energy region, and the other
is the $\Theta$ vacuum \cite{Hoof2}, which is a superposition of the various false
vacua, violating CP symmetry. At lower energies, on the other hand, many nuclear
effective models as the alternatives of QCD were proposed but their applications are
not complete and limited to a few aspects. It is thus the motivation of quantum
nucleardynamics (QND), which is derived from QCD as the consequence of the confinement
and $\Theta$ vacuum, to explain diverse nuclear phenomena at lower energies
consistently. For examples, nuclear issues to be clarified are as follows: Lande's
spin g-factor for nucleon, constant nucleon density, intrinsic quantum number,
nucleon-nucleon scattering data, baryon number conservation, proton number
conservation, etc. These phenomena at relatively lower energies may be partly explained
by nuclear effective models but each effective model is only applicable to limited
issues. In this context, it is proposed that QND as an $SU(2)_N \times U(1)_Z$ gauge
theory \cite{Roh3,Roh31} is the theory for strong interactions of nucleons just as the
Glashow-Weinberg-Salam (GWS) model \cite{Glas} as an $SU(2)_L \times U(1)_Y$ gauge
theory is the theory for electroweak interactions of quarks and leptons.

QND is generated from QCD through dynamical spontaneous symmetry breaking (DSSB)
mechanism, whose details are explained in reference \cite{Roh3}: $SU(3)_C \rightarrow
SU(2)_N \times U(1)_Z \rightarrow U(1)_f$. The Lagrangian density \cite{Roh3,Roh31} of
QND as an $SU(2)_N \times U(1)_Z$ gauge theory has the similar form with QCD as an
$SU(3)_C$ gauge theory without the explicit mass term:
\begin{equation}
{\cal L}_{QND} = - \frac{1}{2} Tr  G_{\mu \nu} G^{\mu \nu}
+ \sum_{i=1}  \bar \psi_i i \gamma^\mu D_\mu \psi_i  + \Theta \frac{g_n^2}{16 \pi^2} Tr G^{\mu \nu} \tilde G_{\mu \nu},
\end{equation}
where the bare $\Theta$ term \cite{Hoof2} is a nonperturbative term added to the
perturbative Lagrangian density with an $SU(2)_N \times U(1)_Z$ gauge invariance.
The subscript $i$ stands for the classes of pointlike spinor $\psi$ and $A_{\mu} = \sum_{a=0} A^a_{\mu}
\lambda^a /2$ stand for gauge fields. The field strength tensor is given by $G_{\mu
\nu} = \partial_\mu A_\nu - \partial_\nu A_\mu - i g_n [A_\mu, A_\nu]$ and $\tilde G_{\mu
\nu}$ is the dual field strength tensor. The $\Theta$ term
apparently odd under both P, T, C, and CP operation.
The proton and neutron as spinors possess up and down colorspins as a
doublet just like up and down strong isospins:
\begin{equation}
{\uparrow \choose \downarrow}_c, \ \uparrow = {1 \choose 0}_c, \ \downarrow = {0
\choose 1}_c .
\end{equation}
This implies that conventional, global $SU(2)$ strong isospin symmetry introduced by Heisenberg \cite{Heis}
is postulated as the combination of local $SU(2)$ colorspin and local $SU(2)$ weak isospin symmetries.
The overall wave function for a nucleon may thus be expressed by
$\psi_N = \psi (\textup{colorspin}) \psi (\textup{isospin}) \psi (\textup{spin}) \psi (\textup{space})$.
There exist two types of nucleons, one of which is the color doublet, which is governed by QND,
and the other of which is color singlet
just as there are two type of quarks and leptons, weak isospin doublet and singlet, in the GWS model:
this concept is not contradicted with the conventional concept for hadrons as color singlet.
In this scheme, gluon is a massive gauge boson rather than a massless gauge boson.
The new concepts of nucleons as color doublets and massive gauge bosons are
required to explain the confinement of quarks, the $\Theta$ vacuum, and the violation of discrete symmetries.
The Yukawa potential due to massive gluon confines quarks and gluons.
Discrete symmetries are nonpertubatively broken by the $\Theta$ vacuum as illustrated in the spectra of baryons:
the CP and T violation of the neutron electric dipole moment with $\Theta \leq 10^{-9}$ \cite{Alta},
the C and CP violation of the baryon asymmetry $\delta_B \simeq 10^{-10}$ \cite{Stei0},
the P violation of no parity partners in baryons and mesons, etc.
It is also confirmed that nucleons conserve the vector current but do not conserve the axial vector current
just as quarks and leptons conserve the (V - A) current but do not conserve the (V + A) current.
The nuclear coupling constant $g_n^2 = c_f g_s^2 = \sin^2 \theta_R g_s^2 = g_s^2/4$ is given in terms of the strong coupling constant $g_s$ and
the color factor $c_f$.
The effective strong coupling constant $G_R/\sqrt{2}= g_n^2/8
M_G^2 \approx 10 \ \textup{GeV}^{-2}$ like Fermi weak constant $G_F/\sqrt{2} = g_w^2/8
M_W^2 \approx10^{-5} \ \textup{GeV}^{-2}$ and the color mixing angle $\sin^2 \theta_R
= 1/4$ like the Weinberg mixing angle $\sin^2 \theta_W = 1/4$ thus play important
roles in nuclear interactions \cite{Roh3}. In the following, it is briefly shown how QND is applied
to strong nuclear interactions at both low and high energies:
the details of QND and
its applications including nucleon mass generation are discussed in two references \cite{Roh3,Roh31}.

There are several, explicit examples about lifetimes and cross sections, supporting
QND as an $SU(2)_N \times U(1)_Z$ gauge theory in analogy
with the GWS model as an $SU(2)_L \times U(1)_Y$ gauge theory.
The state $\Sigma^0$ is formed as a resonance of central mass $1385$ MeV in a $K^{-} p$ interaction:
\begin{math}
K^- + p \rightarrow \Sigma^0 \rightarrow \Lambda + \pi^0
\end{math}
where the Q-value in the decay $130$ MeV and the lifetime $\tau = 1/\Gamma \approx 10^{-23}$ s are estimated from the measured
decay width $\Gamma = 36$ MeV.
If the decay rate $\Gamma \simeq G_R^2 m_\Sigma^5$ is used in analogy with the muon decay rate
$\Gamma = G_F^2 m_\mu^5/192 \pi^3$ of weak interactions, the
$\Sigma^0$-hyperon lifetime becomes the order of $\tau \approx 10^{-23}$ s
compared with the muon lifetime in the order of $10^{-6}$ s.
In the strong decay process, the exchange of massive gluon is taken into account
like the exchange of massive intermediate vector boson in the weak decay process.
The strong decay $\Delta \rightarrow n + p$ and the weak decay
$\Sigma^{\pm} \rightarrow n + \pi^{\pm}$ have almost same $0.12$ GeV kinetic energy, approximately.
Their lifetime ratio becomes
\begin{equation}
\frac{\tau(\Delta \rightarrow n + p)}{\tau(\Sigma \rightarrow n + \pi)}
\simeq \frac{1/G_R^2 m_\Delta^5}{1/G_F^2 m_\Sigma^5}
\simeq \frac{10^{-23} \ \textup{s}}{10^{-10} \ \textup{s}} .
\end{equation}
Similarly, the lifetime ratio between
$\Sigma^0 \rightarrow \Lambda + \pi^0$ and $\Sigma^- \rightarrow n + \pi^-$
is found to be
\begin{math}
\frac{\tau(\Sigma^0 \rightarrow \Lambda + \pi^0)}{\tau(\Sigma^- \rightarrow n + \pi^-)}
\simeq \frac{1/G_R^2 m_{\Sigma^0}^5}{1/G_F^2 m_{\Sigma^-}^5}
\simeq \frac{10^{-23} \ \textup{s}}{10^{-10} \ \textup{s}} .
\end{math}
Since the typical cross section in weak interactions can be estimated by $\sigma \simeq G_F^2 T^2 \approx 10^{-44} \ \textup{m}^2$,
the typical cross section in strong interactions is $\sigma \simeq G_R^2 T^2 \approx 10^{-30} \ \textup{m}^2$, which gives good agreement with experiment result.
In the decay of $\Delta^{++}$, which hints the color quantum number,
$\Delta^{++} \rightarrow \pi^+ + p$, the lifetime of $\Delta^{++}$ is $10^{-23}$ s.
This can be interpreted by the distance of about $1$ fm estimated by the
gluon mass of about $300$ MeV in this scheme.
In the strong interaction of $\pi + p \rightarrow \pi + p$ and weak
interactions of $\nu + p \rightarrow \nu + p$, the cross section ratio
around the kinetic energy $T = 1$ GeV becomes
\begin{math}
\frac{\sigma(\pi + p \rightarrow \pi + p)}{\sigma(\nu + p \rightarrow \nu + p)}
\simeq \frac{G_R^2 T^2}{G_F^2 T^2}
\simeq \frac{10 \ \textup{mb}}{10^{-11} \ \textup{mb}} .
\end{math}
These examples described above explicitly exhibit the typical lifetimes and cross sections in strong interactions as expected.

The color $SU(3)_C$ symmetry generates the $SU(2)_N \times U(1)_Z$ symmetry, which
governs nuclear strong dynamics, and the $U(1)_f$ symmetry, which governs nuclear
electromagnetic dynamics, with the condensation of singlet gluons. The $SU(2)_N \times
U(1)_Z$ symmetry and $U(1)_f$ symmetry using the symmetric color factors, $c^s_f =
(c_f^b, c_f^n, c_f^z, c_f^f) = (1/3, 1/4, 1/12, 1/16)$, are applied to the typical
strong interactions \cite{Roh3}. The color factors are related to the strong color
mixing angle $\sin^2 \theta_R$ just like the isospin factors are related to the
Weinberg mixing angle $\sin^2 \theta_W$: for examples, $c_f^n = \sin^2 \theta_R = 1/4$
and $i_f^w = \sin^2 \theta_W = 1/4$. The factors described above are the pure color
factors due to color charges but the effective color factors used in nuclear dynamics
must be multiplied by the isospin factor $i_f^w = \sin^2 \theta_W = 1/4$ with the weak
Weinberg angle $\sin \theta_W$ since the proton and neutron are an isospin doublet as
well as a color doublet:
\begin{equation} c_f^{eff} = i_f^w c_f = i_f^w (c_f^b, c_f^n, c_f^z, c_f^f)
= (1/12, 1/16, 1/48, 1/64)
\end{equation}
for symmetric configurations. For example, the electromagnetic color factor for the
$U(1)_f$ gauge theory becomes $\alpha_f^{eff} = \alpha_s/64 \simeq 1/137$ when
$\alpha_s = 0.48$ at the QCD scale \cite{Hinc}. The coupling constant $\alpha_f =
\alpha_s/16$ for a $U(1)_f$ gauge theory at the strong scale is used to evaluate
excitation levels. The pure color coupling constant $\alpha_f = \alpha_s/16 =
\alpha_w/4$ has about four times stronger than the coupling constant $\alpha_e =
\alpha_w/4$ for a $U(1)_e$ gauge theory at the weak scale: the effective coupling
constant $\alpha_f^{eff} = i^w_f \alpha_f = \alpha_s/64 = \alpha_e$. The emission of
energetic photons as gamma radiation is typical for a nucleus deexcitation from some
high lying excited state to the ground state configuration. This represents a
reordering of the nucleon in the nucleus with a lowering of mass from the excited mass
to the lowest mass. Electromagnetic transition due to the proton charge and gamma
decay are well established subjects. For another example, nuclei with closed shell
plus one valence nucleon is considered in analogy with the hydrogen atom. The Coulomb
potential originated from color charges is thus automatically realized in
nucleon-nucleon electromagnetic interactions in terms of the $U(1)_f$ gauge theory in
addition to the Coulomb potential originated from isospin electric charges. The
separation energy of a valence nucleon is $m_n \alpha_f^2/2 \simeq 0.42$ MeV with the
coupling constant $\alpha_f \simeq 0.03$.

The conservation of the proton number is the result of the
$U(1)_f$ local gauge theory just as the conservation of the
electron number is the result of the $U(1)_e$ local gauge theory.
The charge quantization is given by $\hat Q_f = \hat C_{3} + \hat
Z_c/2$ where $\hat C_3$ is the third component of the colorspin
operator $\hat C$ and $\hat Z_c$ is the hyper-color charge
operator \cite{Roh31}. This form has the analogy with the electric charge
quantization $\hat Q_e = \hat T^w_{3} + \hat Y^w/2$ \cite{Glas} in
the GWS model and $\hat Q_e = \hat T^s_{3} + \hat Y^s/2$
\cite{Gell} in the quark model. The hyper-color charge operator
may be defined by $\hat Z_c = \hat B + \hat S$ with the baryon
number operator $\hat B$ and the strangeness number operator $\hat
S$ just as the hypercharge operator is defined by $\hat Y = \hat B
- \hat L$ with the baryon number operator $\hat B$ and the lepton
number operator $\hat L$. Color charge quantum numbers for the
proton and neutron are shown in Table \ref{coqu} where the
subscript $d$ denotes the colorspin doublet and the subscript $s$
denotes the colorspin singlet. Nucleons as the color spin doublet
are governed by the $SU(2)_N \times U(1)_Z$ gauge theory just as
leptons or quarks as the isospin doublet are governed by the
$SU(2)_L \times U(1)_Y$ gauge theory in weak interactions. The
conservation of the baryon number is the consequence of the
$U(1)_Z$ local gauge theory just as the conservation of the lepton
number is the consequence of the $U(1)_Y$ local gauge theory \cite{Roh31}.
Baryons are conserved as the colorspin
doublet but are not conserved as the color singlet; this is
analogous to the conservation of leptons as the isospin doublet
but the nonconservation of leptons as the isospin singlet in weak
interactions. The immediate result of the proton number
conservation or the baryon number conservation is shown in the
mass density and charge density of nuclear matter. The effective
charge unit of the charge operator $\hat Q_f$ is $q^{eff}_f =
\sqrt{\pi \alpha_s/64}$ while the electric charge unit of the
charge operator $\hat Q_e$ is $e = \sqrt{\pi \alpha_i/4}$: the
absolute magnitude of $q^{eff}_f$ is the same with that of $e$
since $\alpha_s \simeq 0.48$ at the strong scale and $\alpha_i
\simeq 0.12$ at the weak scale.

Nuclear matter is quantized by the maximum wavevector mode
$N_F \approx 10^{26}$ in one dimension and the total baryon number
$B = N_B = 4 \pi N_F^3/3  \approx 10^{78}$
as the consequence of the baryon number conservation or the baryon
asymmetry $\delta_B \simeq 10^{-10}$ \cite{Stei0}.
Baryon matter quantization is consistent with the
nuclear number density $n_n = n_B = A/(4 \pi r^3/3) \approx 1.95 \times 10^{38} \ \textup{cm}^{-3}$
with the nuclear mass number $A$ and the nuclear matter radius $r$ at the strong scale
$M_G \simeq 10^{-1}$ GeV and is consistent with
Avogadro's number $N_A = 6.02 \times 10^{23} \ \textup{mol}^{-1} \approx 10^{19} \ \textup{cm}^{-3}$
at the atomic scale $M_G \simeq 10^{-8}$ GeV:
the nuclear matter density is comparable to massive gluon density $M_G^3 \simeq 10^{38} \ \textup{cm}^{-3}$ at the strong scale
and atomic matter density is also comparable to massive gauge boson density $M_G^3 \simeq 10^{19} \ \textup{cm}^{-3}$ at the atomic scale.
This argument is also compatible with
the nuclear number density $n_B = 2 k_F^2/3 \pi^2$ with the Fermi momentum of a free nucleon $k_F = 1.33 \ \textup{fm}^{-1}$ and
the Fermi energy $\epsilon_F = k_F^2/2 m_n \approx 37$ MeV and compatible with the nucleus radius
\begin{equation}
r = r_0 A^{1/3} = r_0 n^2
\end{equation}
where the nucleon mass radius is $r_0 = 1/2 m_n \alpha_z \approx 1.2$ fm with the color degeneracy factor $2$
and the principal quantum number $n = A^{1/6} = B^{1/6}$
in analogy with the Bohr radius $a_B = 1/m_e \alpha_e$ of the hydrogen atom.

The Lande spin g-factors for the intrinsic magnetic dipole moment
are $g_{s}^p = 2 \mu_p/mu_N = 5.59$ for a proton and $g_{s}^n = 2
\mu_n/\mu_N = -3.83$ for a neutron where $\mu_N = e/m_p = 3.15
\times 10^{-17} \ \textup{GeV}/T$ is the nuclear magneton. The
g-factors are different with $g_{s} = 2$ for a pointlike electron
and $g_{s} = 0$ for a pointlike neutral particle. The problem of
the nuclear magnetic dipole moment suggests that contributions
from colorspin and isospin degrees of freedom must be included to
nucleons. The shifted values for the proton and neutron, $3.59$
and $- 3.83$ are almost identical and they mostly come from the
combined contribution of colorspin and isospin. The mass ratio of
the proton and the constituent quark, $m_p/m_q \sim 2.79$, also
represents combined colorspin, isospin, spin degrees of freedom.
The contribution $|g_{s}| \simeq 3.83$ common for $\mu_p$ and
$\mu_n$ comes from the magnetic dipole moment due to the
$SU(2)_N \times U(1)_Z$ symmetry for color charges and the
$SU(2)_L \times U(1)_Y$ symmetry for isospin charges while the
contribution $g_{s} \simeq 1.76$ only for $\mu_p$ might come from
the magnetic dipole moment due to the $U(1)_f$ gauge symmetry.
This interpretation is justified if the coupling constant $g_f =
\sqrt{c_f^f \alpha_s} = \sqrt{\alpha_s/16} \simeq 1.7 \ e$ for the
$U(1)_f$ gauge theory and the coupling constant $g_b = \sqrt{c_f^b
\alpha_s} = \sqrt{\alpha_s/3} \simeq 3.8 \ e$ for the $SU(2)_N
\times U(1)_Z$ gauge theory. The description above reflects the
mixed contribution of colorspin, isospin, and spin degrees of
freedom and the total angular momentum
\begin{equation}
\vec J = \vec L + \vec C + \vec I + \vec S
\end{equation}
and the extension of $\vec J = \vec L + \vec S$.

Quantum numbers of nucleon-nucleon systems are summarized in Table
\ref{nnqu} when colorspin degrees of freedom are taken into
account in addition to isospin and spin degrees of freedom. Cross
sections for the nucleon-nucleon (NN) scattering as an $SU(2)_N$
gauge theory in terms of massive gluon exchange show excellent
agreement with measurement data.  According to conventional strong isospin
invariance, three types of scattering such as the nn, pp, and pn
scattering with strong isospin one (spin zero) exhibit almost the
same cross sections. The cross section for the nn, pp, or pn
scattering as a colorspin triplet is expressed by
\begin{equation}
\sigma = \frac{4 G_R^2 T^2}{\pi} \frac{1}{1 + 4 T^2/M_G^2}
\end{equation}
in the center of mass energy $T$ since ${\cal M} \sim G_R J_c^\mu
J^\dagger_{c \mu}$ in terms of the effective strong coupling
constant $G_R = \sqrt{2} c^n_f g_s^2/8 M_G^2$. The theoretical cross
section of the nn or pp scattering as an $SU(2)_N$ gauge theory at
high energies about from $2$ GeV to $10^{3}$ GeV is saturated to
the experimental one of about $40$ mb \cite{Galb} and the cross section at low energies
from $0.6$ GeV to $2$ GeV is roughly proportional to $T^2$: $c^n_f =
1/4$, $\alpha_s = 0.48$, $M_G \approx 300$ MeV, and $G_R = 10 \
\text{GeV}^{-2}$ are used in this evaluation. The symmetric
$SU(2)_N$ colorspin interaction for the isospin triplet and spin
singlet contribution is commonly involved in the above three types
of nucleon-nucleon interactions with the massive gluon exchange.

The cross section in strong interactions as a $U(1)_Z$ gauge
theory at relatively low energies can be nonrelativistically obtained
using the Yukawa potential $V(r) = \sqrt{c^z_f \alpha_s} e^{- M_G
(r - l_{QCD})}/r$:
\begin{equation}
\sigma = 4 \pi \frac{(c^z_f \alpha_s m_n)^2}{M_G^4} \frac{1}{1 + 4 m_n T/M_G^2}
\end{equation}
where $c^z_f = - 1/6$ (or $1/12$) is the asymmetric (symmetric) color factor, $m_n =
0.94$ MeV is the nucleon mass, $T$ is the incident particle energy, and the gauge
boson mass $M_G \approx 140$ MeV. The cross section data in the region below the
energy $0.3$ GeV or above the range $1.5$ fm are obtained by using the above formula,
which is definitely dependent on angular momenta. The calculated cross section data in
the limit $T \rightarrow 0$ give agreement with observed data for cross sections
($\sigma = 4 \pi a^2$) $\sigma_{pp} \simeq \sigma_{nn} \simeq 35$ b and $\sigma_{pn}
\simeq 66$ b for spin singlet and $\sigma_{pn} \simeq 4$ b for spin triplet
\cite{Wils}. The effective (running) coupling constant $\alpha_z = c_f^z \alpha_s =
\alpha_s/16$ becomes stronger at lower energies: $\alpha_z =c_f^z \alpha_s \simeq
0.03$ at about $T=100$ MeV. The cross section difference in strong isospin triplet is
mainly due to the contribution of colorspins: $\sigma_{pn}$ has contributions from
both colorspin triplet ($c=1$) and colorspin singlet ($c = 0$) as shown in Table
\ref{nnqu}. In the strong isospin triplet scattering ($i^s = 1$), almost the same
cross section data are consequences of the similar invariant amplitude magnitude for
color charged current and color neutral current mediated by color charged massive
gluons ($A^\pm$) and neutral massive gluon ($B^0$), respectively, just as shown that
the relative strength of the charged current and neutral current is the same in the
GWS model \cite{Glas}. It is thus emphasized that QND as the $SU(2)_N \times U(1)_Z$
gauge theory produces the cross section data, which do not have divergence problems so
that QND is renormalizable, from the zero energy limit to the order of $10^{2}$ GeV.

In conclusion, QCD produces QND as the $SU(2)_N \times U(1)_Z$ gauge theory for
nuclear interactions and then produces the $U(1)_f$ gauge theory for massless gauge
boson (photon) dynamics. Comparison between QND and effective models is given in Table
\ref{comp}. QND is applicable to various aspects in the wide energy range but
effective models are only effective to few aspects of nuclear phenomena in the rather
small energy range.  The effective strong coupling constant $G_R/\sqrt{2}=
c_f g_s^2/8 M_G^2 \approx 10 \ \textup{GeV}^{-2}$ like Fermi weak constant
$G_F/\sqrt{2} = g_w^2/8 M_W^2 \approx10^{-5} \ \textup{GeV}^{-2}$ are used to study
nuclear interactions. The proton number conservation is the result of the $U(1)_f$
gauge theory and the baryon number conservation ($B = N_B \approx 10^{78}$) is the
consequence of the $U(1)_Z$ gauge theory for strong interactions: the charge
quantization $\hat Q_f = \hat C_3 + \hat Z_c/2$ with the hyper-color operator $\hat
Z_c = \hat B + \hat S$ for the $U(1)_f$ gauge theory. The mass density and charge
density of nuclear matter are the consequence of the proton number conservation or the
baryon number conservation.  The proton and neutron are assigned as a
colorspin plus weak isospin doublet instead of as a strong isospin doublet.
The extension of the total angular momentum $\vec J = \vec L + \vec
S + \vec C + \vec I$ from  $\vec J = \vec L + \vec S$ may explain the Lande spin
g-factors of magnetic dipole moments for the proton and neutron, $g_s^p = 5.59$ and
$g_s^n = - 3.83$, respectively. The cross sections of nucleon-nucleon scattering
(pp, nn, or np) are compatible with QND as an $SU(2)_N \times U(1)_Z$ gauge theory.
More testable predictions or already confirmed predictions from QND are given as
follows: parity violation in meson and baryon spectra, charge conjugation violation in
baryon spectra, time reversal and CP violation in the electric dipole moment for the
neutron ($\Theta \leq 10^{-9}$), the nonconservation of color singet proton and
neutron, the nonconservation of the axial vector current, strong coupling constant hierarchy,
the color mixing angle $\sin^2 \theta_R \simeq 1/4$, the assignment of strong isospin
as colorspin plus weak isospin. Furthermore, QND possessing colorspin degrees of
freedom may be successful in explaining the various nuclear phenomena such as
lifetimes and cross sections of nuclear scattering and reaction, shell model,
meson-nucleon scattering, nuclear potential, nuclear binding energy, gamma ray, etc.
over the wide energy range \cite{Roh3}.

\begin{table}[h]
\caption{\label{coqu} Color Quantum Numbers of Nucleons}
\end{table}
\centerline{
\begin{tabular}{|c|c|c|c|c|} \hline
Baryons & $C$ & $C_3$ & $Z_c$ & $Q_f$ \\ \hline \hline
$p_d$ & $1/2$ & $1/2$ & $1$ & $1$ \\ \hline
$n_d$ & $1/2$ & $-1/2$ & $1$ & $0$ \\ \hline
$p_s$ & $0$ & $0$ & $2$ & $1$ \\ \hline
$n_s$ & $0$ & $0$ & $0$ & $0$ \\ \hline
\end{tabular}
}

\vspace{1cm}

\begin{table}
\caption{\label{nnqu} Quantum Numbers of Nucleon-Nucleon Systems
($i^s$: strong isospin, $i$: weak isospin)}
\end{table}
\centerline{
\begin{tabular}{|c|c|c|} \hline
State & $i^s=1$ & $i^s=0$ \\ \hline \hline
pp & $i=1, s=0, c=1$ &  \\ \hline
nn & $i=1, s=0, c=1$ &  \\ \hline
pn & $i=1, s=0, c=1$ & $i=0, s=1, c=0$ \\
   & $i=1, s=1, c=0$ & $i=0, s=0, c=1$ \\ \hline
\end{tabular}
}

\newpage
\onecolumn

\begin{table}
\caption{\label{comp} Comparison between Quantum Nucleardynamics
and Effective Models}
\end{table}
\centerline{
\begin{tabular}{|c|c|c|} \hline
Classification & QND & Effective Models \\ \hline \hline Exchange Particles & massive
gluons & model dependent \\ \hline DSSB & yes & no \\ \hline Discrete symmetries (P,
C, T, CP) & breaking &  no \\ \hline Confinement & yes & no \\ \hline $\Theta$ vacuum
& yes & no
\\ \hline Baryon number conservation & $U(1)_Z$ gauge theory ($N_B
\simeq 10^{78}$) & unknown \\ \hline Proton number conservation &
$U(1)_f$ gauge theory & unknown \\ \hline Nuclear electromagnetic
interaction & $U(1)_f$ gauge theory & no \\ \hline Intrinsic
angular momenta & colorspin, weak isospin, spin & spin
\\ \hline Hadron mass generation & yes & unknown \\ \hline NN
scattering cross section & $G_R^2 T^2$ (from high to low energy) &
only at low energy \\ \hline Hadron decay time & $1/G_R^2 m_h^5$ &
no \\ \hline Neutron electric dipole moment & $\Theta \simeq
10^{-12}$ & no \\ \hline Free parameters & coupling constant &
many \\ \hline Renormalization & yes & model dependent \\ \hline
\end{tabular}
}

\end{document}